\documentclass{iopart}
\usepackage{graphicx}
\usepackage{dcolumn}
\usepackage{bm}

\newcommand{\gsim}{{\textstyle{\lower 2pt \hbox{$>$} \atop \raise 1pt \hbox{$ \sim$}}}}

\newcommand\onlinecite[1]{\cite{#1}}

\begin{document}

\title{Electronic and atomic shell structure in aluminum nanowires}

\author{A.~I. Mares$^{1}$, D.~F. Urban$^{2}$, J. B\"{u}rki$^{3}$, H. Grabert$^{2}$, \\
        C. A. Stafford$^{3}$ and J.~M. van Ruitenbeek$^{1}$}


\address{$^{1}$Kamerlingh Onnes Laboratorium, Universiteit Leiden,
    P.O. Box 9504, 2300 RA Leiden, The Netherlands}
\address{$^{2}$Physikalisches Institut, Albert-Ludwigs-Universit\"{a}t, D-79104 Freiburg, Germany}
\address{$^{3}$Department of Physics, University of Arizona, Tucson, Arizona 85721, USA}

\ead{urban@physik.uni-freiburg.de}

\submitto{\NT,\hspace{1cm}\today}

\begin{abstract}

We report experiments on aluminum nanowires in ultra-high vacuum
at room temperature that reveal a periodic spectrum of
exceptionally stable structures. Two ``magic'' series of stable
structures are observed: At low conductance, the formation of
stable nanowires is governed by electronic shell effects whereas
for larger contacts atomic packing dominates. The crossover
between the two regimes is found to be smooth. A detailed
comparison of the experimental results to a theoretical stability
analysis indicates that while the main features of the observed
electron-shell structure are similar to those of alkali and noble
metals, a sequence of extremely stable wires plays a unique role
in Aluminum. This series appears isolated in conductance
histograms and can be attributed to ``superdeformed''
non-axisymmetric nanowires.

\end{abstract}

\maketitle

\section{Introduction}

Metal nanowires have attracted considerable interest in the past
decade, experimental research has burgeoned, and various
techniques have been employed to produce them for many different
materials (see\ \onlinecite{Agrait03} for a recent review).
Conductance  histograms, built out of thousands of conductance
traces recorded during the breaking of a nanowire in a
mechanically controllable break junction (MCBJ), reveal the
existence of exceptionally stable wires: Due to their stability,
these wires are formed more often than others, and show up as
peaks in a conductance histogram taken at a temperature large
enough to allow the contact to explore the available
configuration space.

Such experiments on alkali metals
\cite{yanson99,Yanson01,Yanson04} have revealed a sequence of
stable ``magic'' wires that is periodic in $\sqrt{G}$, where $G$
is the electrical conductance. A theoretical analysis of nanowire
stability \cite{kassubek01,urban04,burki05a} within the nanoscale
free-electron model (NFEM)
\cite{stafford97,ruitenbeek97,Buerki05b} has connected this magic
sequence of wires to the electron shell structure, similar to
what is known for metal clusters
\cite{deheer93,brack93,martin96}. Groups of discrete transverse
electronic states are bunched together, with gaps between them.
When the electronic Fermi energy is within one of these gaps, the
wire is exceptionally stable. The periodicity in $\sqrt{G}$ can
be understood semi-classically: writing the mesoscopic
contribution to the wire energy as a series in classical periodic
orbits in a circular billiard
\cite{yanson99,kassubek01,Buerki05b}, the main contribution is
found to come from the diameter orbit, and to be periodic in the
wire radius, and thus proportional to $\sqrt{G}$ via the Sharvin
formula \cite{torres94,hoppler98}.
A supershell structure was also observed \cite{Yanson00}
in the form of a periodic modulation of the peak heights, and it
was argued that the stable wires at the nodes of the supershell
structure are non-axisymmetric \cite{urban04,urban04b,Urban06}.

At larger wire diameters, and therefore larger conductance, a
crossover to an atomic shell structure was found \cite{yanson01a}.
In this case, shell closings correspond to crystalline wires with the
completion of additional atomic layers.
Recently, electron-shell structure has also been observed for the
noble metals gold \cite{Diaz03,mares04}, silver \cite{mares05}, and
copper \cite{mares05}. Except for the latter, the crossover to
atomic shell effects at larger diameter could be
confirmed \cite{mares04,mares05}.

Inspired by these results and by the research done on metal
clusters, we extend our investigation to multivalent metals,
choosing as a first step aluminum.

Conductance histograms for aluminum atomic size contacts recorded
using the MCBJ technique at 4.2 K were first reported by Yanson
\etal \cite{yanson97}. Four peaks were identified situated close
to 1, 2, 3 and 4 times the conductance quantum $G_0=2e^2/h$.
Scheer \etal \cite{scheer97} analyzed the transmission modes for
Al point contacts by recording current-voltage characteristics in
the superconducting state. For a single atom, three modes were
found to contribute to the conductance, with individual
transmissions lower than one summing together to a total
conductance of about one quantum unit.

A previous experiment on aluminum shell structure \cite{medina03}
reports both experimental results and molecular dynamics
simulations of the evolution of Al nanowires while breaking. The
experiment was performed using a scanning tunnelling microscope
(STM) at room temperature (RT) in ultra high vacuum (UHV). The
authors attribute the oscillation period of the peaks to filling
of octagonal facets, claiming that the electronic shell effect is
absent. Our results reported below are significantly different
since we present evidence of electronic shell effects. The
difference with our experiments appear in several aspects:
Firstly, in the histogram from \onlinecite{medina03} there is a
large contribution in the low conductance regime, including below
1 $G_0$. This is usually an indication of the presence of
contaminants, especially expected in the case of STM soft
indentations \cite{hansen00}. Secondly, in \onlinecite{medina03}
the peaks have low weight and a derivative is performed in order
to identify their position. Moreover, from one single histogram
one cannot exclude the occurrence of electronic shell effects,
since electronic and atomic shell effects compete in the
stabilization of the nanowire and in different measurements one
or the other effect can dominate, as it will be shown below.
Although the experimental results in \onlinecite{medina03} agree
with the theoretical predictions based on molecular dynamics
simulations, the latter calculation does not take into account
the electronic contribution to the total energy of the nanowire,
excluding the electronic shell effects from consideration.

This paper is organized as follows. The experimental results are
presented in \sref{sec:ExperimentalResults}, followed by a
detailed analysis of the data in \sref{sec:DataAnalysis}. A
theory explaining the observed electronic shell structure and
allowing the prediction of the shapes of the wires is presented in
\sref{sec:ElectronicShellEffects}. Atomic shells and the
crossover between electronic and  atomic shell effects are
analyzed in \sref{sec:AtomicShellEffects}.
\Sref{sec:Superdeformed} addresses the issue of superdeformed
nanowires. Finally, a short summary is given in \sref{sec:Summary}.

\section{Experimental results}
\label{sec:ExperimentalResults}

\begin{figure}[b]     
  \begin{center}
    \includegraphics[width=8cm]{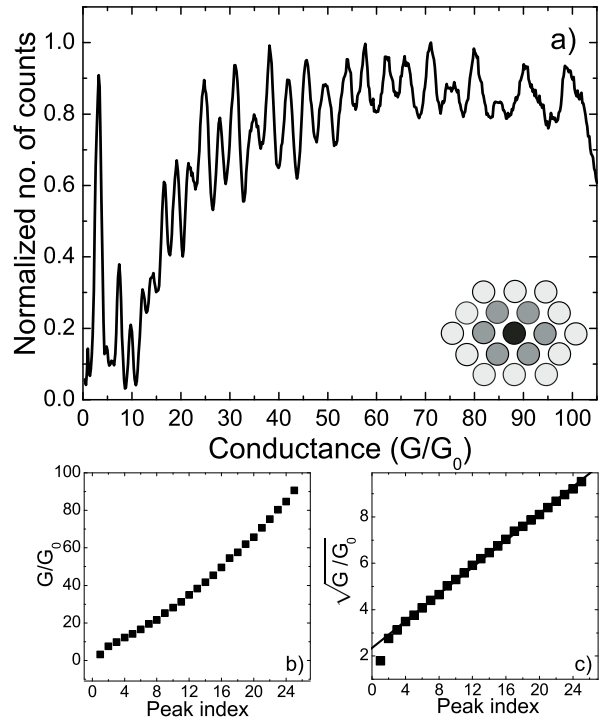}
  \caption{(a) Conductance histogram for aluminum at RT-UHV constructed
    from 7000 scans at a bias voltage of 50 mV. The inset shows the
    presumed atomic positions in the normal cross-section of the wire. In (b) we plot the
    conductance values at the peak position vs.\ their index while in
    (c) $ \sqrt{G/G_0}$ for the peak positions vs.\ peak index. For (c)
    a good linear dependence is obtained with a period $\Delta
    \sqrt{G/G_0}$=0.29 $\pm$0.002.} \label{fig:Al4010}
  \end{center}
\end{figure}

\begin{figure}[b]     
  \begin{center}
     \includegraphics[width=8cm]{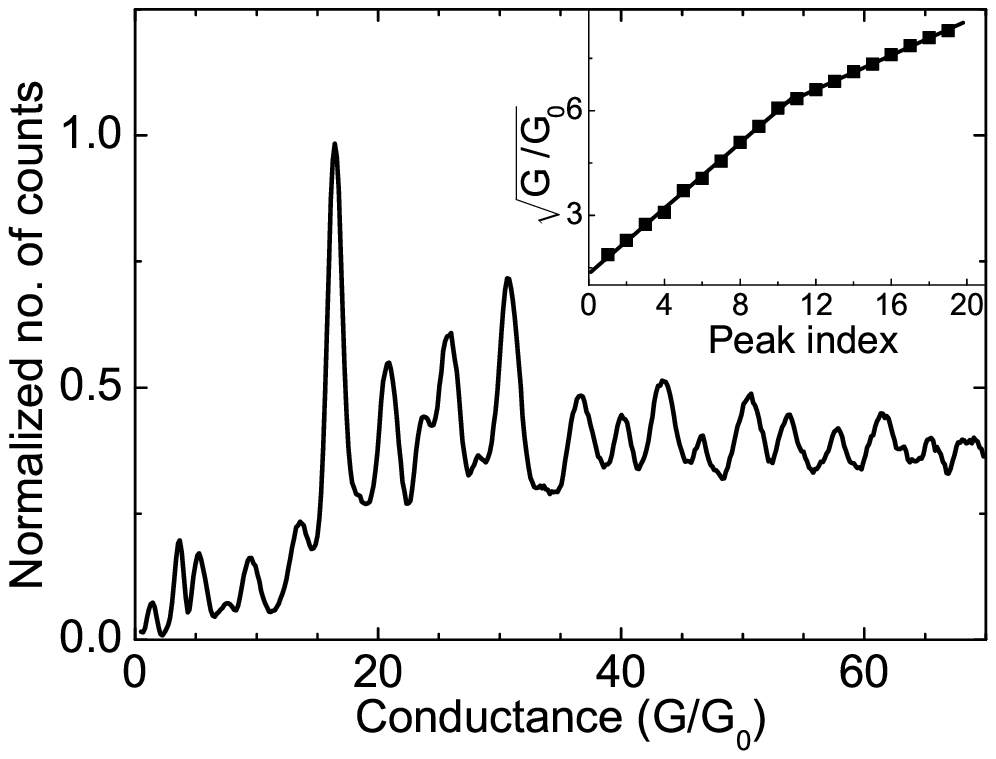}
  \end{center}
  \caption{Conductance histogram for Al taken under RT-UHV conditions, recorded at a
     bias voltage of 20 mV constructed from 2000 scans. Inset: the peak
     oscillation has two periods: in the low conductance regime up to
     about 40 $G_{0}$ the period is $\Delta \sqrt{G/G_0}$= 0.47 $\pm$0.03 whereas for
     larger diameters we find $\Delta \sqrt{G/G_0}$= 0.25 $\pm$0.01.}
     \label{fig:G22}
\end{figure}

\begin{figure}     
  \begin{center}
     \includegraphics[width=8cm]{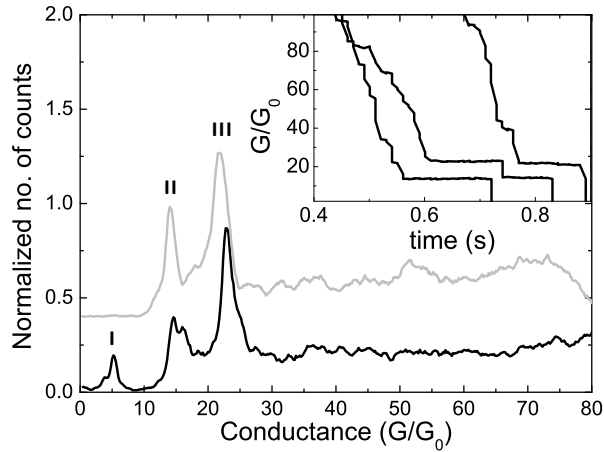}
  \end{center}
     \caption{Special type of histograms recorded for Al
     at RT-UHV showing two dominant peaks near 14 and 22
     $G_{0}$, while the lower curve shows a third peak near 5 $G_{0}$.
     The two curves are vertically offset for clarity.
     The histograms are recorded on two different samples,
     both at a bias voltage of 20 mV containing 1000 traces each.
     In the individual conductance traces long
     plateaus can be seen at these values, shown in the inset.}
     \label{fig:histSuperDeformed}
\end{figure}

We report conductance histograms recorded for Al nanowires in UHV
at room temperature using the MCBJ set-up presented in
\onlinecite{mares04} and \onlinecite{mares05}. A metallic nanowire
is bent in a controlled way and its diameter at the weakest point
is reduced down to atomic size. The wire is initially broken under
UHV, ensuring in this way clean surfaces. In order to perform a
reliable statistical analysis we have investigated six different
samples. The behavior was reproduced on every sample and typical
histograms are shown in figures \ref{fig:Al4010}, \ref{fig:G22},
\ref{fig:histSuperDeformed}. In the histogram from
\fref{fig:Al4010} we can see a remarkably long series of peaks up
to a conductance value of 100 $G_{0}$.

The first step in finding the origin of the peaks in the
conductance histogram is to verify whether the peaks show a
certain periodicity. We can see in \fref{fig:Al4010} (b) and (c)
that a linear dependence is found when plotting the square root
of the conductance as a function of the peak index, while when
plotting directly the conductance the dependence is far from
linear. Note that the square root of the conductance is
proportional to the root-mean-square (rms) radius $\rho$ of the
wire, as can be inferred from the corrected Sharvin formula
\cite{torres94,hoppler98},
\begin{eqnarray}
\label{eq:SharvinConductance}
    G_S&=&G_0\left(\frac{k_F^2\rho^2}{4}-\frac{k_F{\cal
    P}}{4\pi}+\frac{1}{6}\right),
\end{eqnarray}
which gives a semi-classical approximation to the conductance.
Here $\cal P$ is the perimeter of the wire's cross section and $k_F$ is the Fermi
wavevector. In the case of the histogram of \fref{fig:Al4010}
the period found is $\Delta \sqrt{G/G_0}$=0.29 $\pm$0.002.

As for the other metals, we have studied how the histograms vary
from one experiment to the next, and a characteristically
different histogram can be seen in \fref{fig:G22}. There a
crossover to a different period occurs. Two periods of oscillation
can be identified around $\Delta \sqrt{G/G_0}$=0.47 and 0.25, with
a crossover at a conductance value of about 40 $G_{0}$. The first
period is similar to the period of the electronic shell effect in
noble metals \cite{mares04,mares05}. The second one, similar to
the period of the histogram of \fref{fig:Al4010}, is somewhat
larger than what was obtained for the atomic shell effect in noble
metals, for which typically $\Delta \sqrt{G/G_0}$ is found to be
around 0.21.

Sometimes a new type of histogram appears having only two or
three peaks as one can see in \fref{fig:histSuperDeformed}. This
type of histogram appears at the beginning of the measurements on
a fresh sample just after the initial breaking, and we see it in 7
out of 30 histograms. The position of the two major peaks is
always close to 14 and 22 $G_{0}$. In two of these histograms a
peak near 5 $G_{0}$ also appears. All three peaks are related to
very long plateaus at these values in the individual conductance
traces [see inset of \fref{fig:histSuperDeformed}].

\begin{figure}[t]       
  \begin{center}
    \includegraphics[width=8cm]{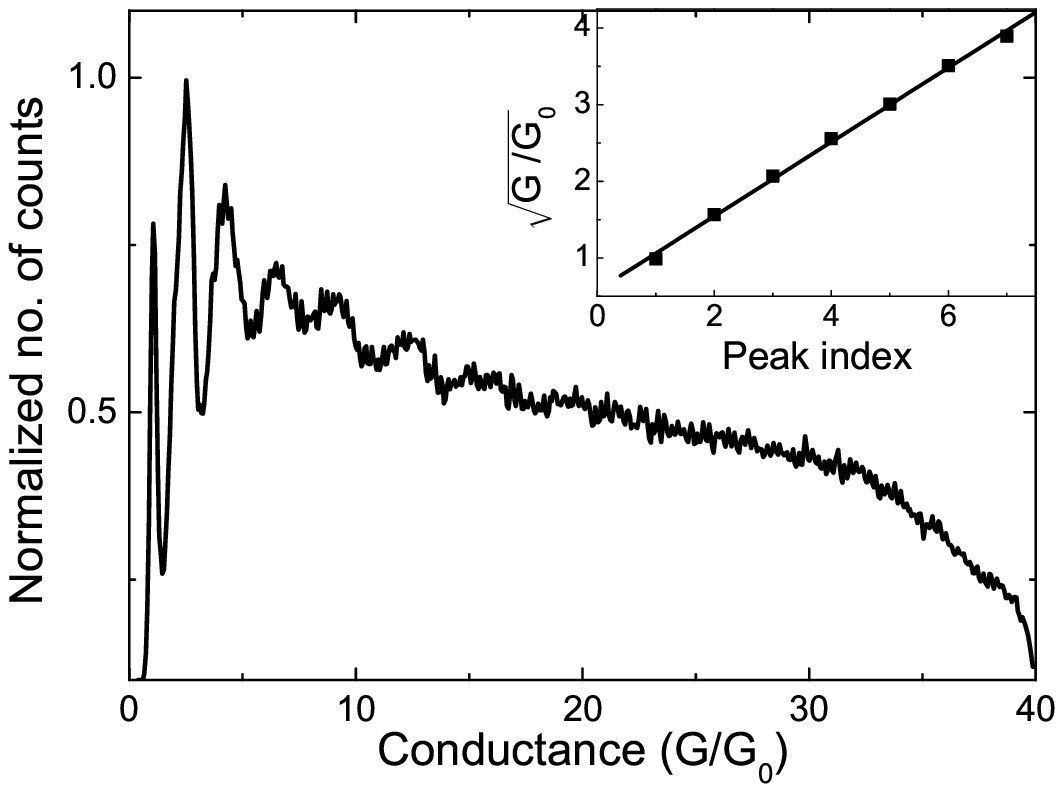}
  \end{center}
  \caption{Conductance
    histogram for Al at room temperature in air constructed from 1000
    traces at a bias voltage of 20 mV. Inset: the peak oscillation
    has a period  $\Delta \sqrt{G/G_{0}}$=0.48 $\pm 0.04$.}
  \label{fig:GAlair}
\end{figure}

Finally, in \fref{fig:GAlair} we show a conductance histogram
recorded for aluminum in air at room temperature. We see distinct
peaks that have a period close to the electronic shell effect
found in noble metals and for Al in UHV. It is known that bulk Al
is a very reactive metal that forms an oxide layer on the surface.
Possibly this layer prevents the immediate contamination inside
the bulk of the nanowire.

\section{Theoretical analysis of Al histograms}\label{sec:DataAnalysis}

\begin{figure}[t]       
  \begin{center}
     \includegraphics[width=8cm, angle=-90]{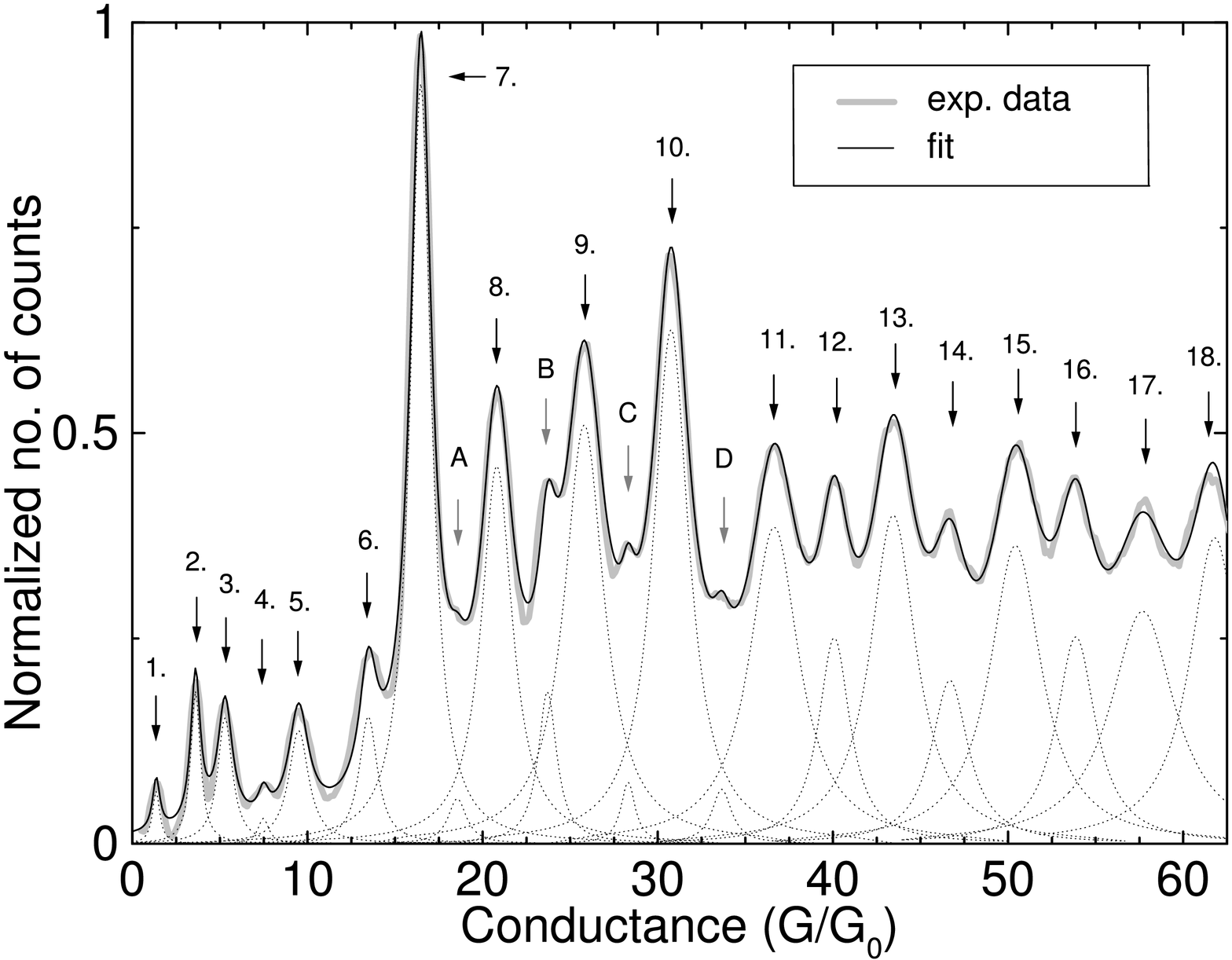}
  \end{center}
  \caption[]{Experimental conductance histogram data from \fref{fig:G22}
    fitted with lorentzian peaks.  The dotted lines show the individual peaks.
    The positions, weights, and widths of the peaks are given in \tref{tab:LorentzFit}.}
  \label{fig:LorentzFit}
\end{figure}

In order to obtain more information about the stable wire
configurations realized in experiments, we have analyzed in
detail the conductance histogram of \fref{fig:G22}. We fit the
data set with a function consisting of a sum of peaks of
lorentzian shape with height, width and peak position as fitting
parameters, and the result is shown in \fref{fig:LorentzFit} and
\tref{tab:LorentzFit}. Note that the number of peaks used for the
fit is not unique, as the fit can always be improved by including
additional broad peaks of lower intensities. However, the
parameters of additional peaks can not be determined with high
confidence, especially at larger conductance where the curve is
more noisy.

Besides the 18 dominant conductance peaks, marked with numbered
black arrows in \fref{fig:LorentzFit}, we can identify a
number of very shallow peaks in the conductance histogram. They
originate from less stable wire configurations, that are less
frequently realized. These ``peaks'' are marked with capital
letters A--D in \fref{fig:LorentzFit}.

\Fref{fig:Periodicity} shows $\sqrt{G}$ versus peak number for
three different peak-selection criteria: Peaks 1--18 of
\tref{tab:LorentzFit} are plotted as red squares, reproducing the
inset of \fref{fig:G22}, with a transition from electron-shell to
atomic-shell structure at a conductance $G\sim 40 G_0$.  When the
shallow peaks A--D are included (green stars), the transition to
atomic-shell structure occurs already at $G\sim 20 G_0$.
Conversely, when only the largest-weight peaks ($>5\%$) at high
conductance are included (blue $+$ symbols), the slope
corresponding to electron-shell structure is seen to extend all
the way up to the highest peak fit, $G\sim 60 G_0$.  Note that at
a temperature sufficient to display electron-shell effects, the
weights of the peaks with $G< 20 G_0$ may be suppressed due to
thermally-induced breaking of the contact, as previously observed
in alkali metals \cite{yanson99}. \Fref{fig:Periodicity} suggests
that electronic- and atomic-shell effects compete in the regime
$20 G_0 < G < 60 G_0$, with atomic-shell effects becoming
increasingly important with increasing $G$.

\Table{\label{tab:LorentzFit} Conductance $G_{\rm exp}$, relative
weight, and width $\Delta G$ of the peaks in the histogram shown
in \fref{fig:LorentzFit}. The values were obtained by fitting the
data set to a sum of lorentzian peaks, with height, width, and
peak position as fitting parameters. The next two columns give the
parameters of the stable Al nanowires determined by a linear
stability analysis within the NFEM: $G_{th}$ is the quantized
conductance and $\lambda$ is the quadrupole deformation [cf.\
\eref{eq:quadrupole}]. The last two columns give the mean channel
transmission $\langle T_n \rangle\equiv G_{exp}/G_{th}$ or series
resistance $R_s$ needed to shift the theoretical peaks down to the
corresponding experimental values. For completeness, the NFEM
predictions are shown for peaks 1--5, although the model is not
expected to be quantitatively accurate for few-atom contacts in
multivalent metals, as evidenced by the large values of the
effective series resistance, or the small channel transmissions
\cite{scheer97}. The last three lines of the Table give the
positions of the three main peaks of \fref{fig:histSuperDeformed},
and the corresponding theoretical parameters.}
    \br
    peak&
     $\begin{array}{cc} {G}_{\rm{exp}} \\ \left[G_0\right]\end{array}$ &
     $\begin{array}{cc} {\rm{weight}} \\ \left[\%\right]\end{array}$ &
     $\begin{array}{cc} \Delta{G} \\ \left[G_0\right]\end{array}$ &
     $\begin{array}{cc} {G_{th}} \\ \left[G_0\right]\end{array}$ &
     $\lambda$ &
     $\langle T_n \rangle$ &
     $\begin{array}{cc} {R_s} \\ \left[\Omega\right]\end{array}$
    \\
    \mr
  1  &  1.4 & 0.2 & 0.5  & 3  & 0    & 0.45 & 5185 \\
  2  &  3.6 & 0.8 & 0.7  & 5  & 0.17 & 0.72 & 993 \\
  3  &  5.3 & 1.1 & 1.0  & 6--7 & 0--0.3 & 0.76--0.88 & 289--596 \\
  4  &  7.5 & 0.2 & 0.7  & 9   & 0.11 & 0.83 & 291 \\
  5  &  9.5 & 1.3 & 1.4  & 12  & 0   & 0.79 & 283 \\
    \mr
  6  & 13.5 & 1.3 & 1.3  & 14 & 0.3 & 0.96 & 36 \\
  7  & 16.5 & 9.2 & 1.5  & 17 & 0    & 0.97 & 24 \\
  A  & 18.6 & 0.5 & 1.3  & -- & --   & -- & -- \\
  8  & 20.8 & 6.7 & 2.2  & 23 & 0    & 0.90 & 59 \\
  B  & 23.7 & 1.6 & 1.3  & -- & --   & -- & -- \\
  9  & 25.8 & 9.5 & 2.8  & 29 & 0.06 & 0.89 & 55 \\
  C  & 28.3 & 0.5 & 1.0  & -- & --   & -- & -- \\
  10 & 30.8 & 10.5 & 2.5 & 34 & 0    & 0.90 & 40 \\
  D  & 33.7 & 0.6 & 1.4  & -- & --   & -- & -- \\
  11 & 36.7 & 9.0 & 3.4  & 42 & 0    & 0.87 & 45 \\
  12 & 40.0 & 3.7 & 2.2  & -- & --   & -- & -- \\
  13 & 43.4 & 8.7 & 3.2  & 51 & 0    & 0.85 & 44 \\
  14 & 46.7 & 3.5 & 2.6  & -- & --   & -- & -- \\
  15 & 50.4 & 8.9 & 3.6  & 59 & 0.05 & 0.85 & 37 \\
  16 & 53.9 & 4.7 & 2.7  & -- & --   & -- & -- \\
  17 & 57.7 & 8.6 & 4.5  & 67 & 0    & 0.86 & 31 \\
  18 & 61.8 & 8.9 & 3.5  & 72 & 0.04 & 0.86 & 30 \\
  \mr
  I  &  5.3 & --  & --   & 6--7 & 0.3 &  0.76--0.88 & 289--596 \\
  II  & 14.1 & -- & -- & 14 & 0.3 & 1.0 & 0 \\
  III  & 21.9 & -- & -- & 23 & 0.3 & 0.95 & 28 \\
  \br
 \endTable

\subsection{Electronic Shell Effects}
\label{sec:ElectronicShellEffects}

Recently, the stability of alkali and noble metal nanowires was
extensively studied theoretically
\cite{kassubek01,urban04,burki05a,urban04b,Urban06,Urban03,Buerki03}
using the Nanoscale Free-Electron Model (NFEM)
\cite{stafford97,Buerki05b}. It was found that nanowire stability
is determined by a competition between surface and shell
contributions. A linear stability analysis considering
cylindrical geometries as well as wires with broken axial
symmetry revealed a ``magic'' sequence of stable nanowires, that
allows for a consistent interpretation of experimental
conductance histograms for alkali metals, including both the
electronic shell and supershell structures
\cite{urban04,urban04b}. The same sequence of stable wires is
observed in histograms for Au \cite{Diaz03,mares04}, suggesting a
possible universality.

\begin{figure}[t]      
  \begin{center}
    \includegraphics[angle=270, width=0.6\textwidth]{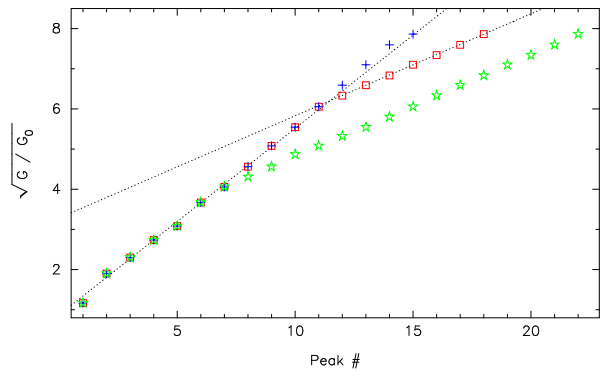}
  \end{center}
  \caption{(colour online) Linear dependence of the peak positions in $\sqrt{G/G_{0}}$ using
    various selection criteria:
    (i) Main peaks, as in the inset of \fref{fig:G22} (red squares);
    (ii) Keeping only the peaks with weight larger than 5\% at large
    conductance (blue $+$ symbols);
    (iii) Keeping all peaks in \tref{tab:LorentzFit} (green stars).
    }
  \label{fig:Periodicity}
\end{figure}

Guided by the importance of conduction electrons in the cohesion
of metals, and by the success of the jellium model in describing
metal clusters \cite{deheer93,brack93}, the NFEM replaces the
metal ions by a uniform, positively charged background that
provides a confining potential for the electrons. The electron
motion is free along the wire, and confined in the transverse
directions. The surface properties of various metals can be fit
by using appropriate surface boundary conditions
\cite{Garcia-martin96,urban04}.

Aluminum is a trivalent metal, but the Fermi surface of bulk Al
resembles a free-electron Fermi sphere in the extended-zone
scheme. This suggests that the NFEM might be applicable to Al
nanowires, although the continuum approximation is more severe
than for monovalent metals. A first-principles theoretical
calculation of the electronic structure of Al nanowires indeed
reveals that, although the atomic structure has to be taken into
account, the band structure has a free-electron like dispersion
relation \cite{ditolla00}. Free-electron jellium
models \cite{brack93} have already successfully been used for
modeling Al clusters \cite{Heer89,Milani91,Jarrold93} and
explaining the occurrence of magic
numbers \cite{Schriver90,Persson91,Pellarin93}. De Heer \etal
\cite{Heer89,Milani91} have specifically addressed the
question of the applicability of the jellium model by comparing
measured electronic properties of Al$_n$ clusters to the values
calculated with the jellium model. For the polarizability they
find good agreement \cite{Heer89} for $n>40$, while when
considering ionization potentials and electron affinities, it is
found that clusters with $n>13$ are already accurately described
by the jellium model \cite{Milani91}.
Using a charge neutrality argument, the atomic volume is given by
$v_a = 9\pi^2/k_F^3$ for a trivalent metal, the radius $R$ of an
Al$_{13}$ cluster can be estimated as $k_F R = 6.5$, while that of
Al$_{40}$ is $k_F R = 9.5$. Aluminum nanowires with these radii
would have electrical conductances of $7~G_0$ and $18~G_0$,
respectively, according to \eref{eq:SharvinConductance}. We
therefore expect the NFEM to become quantitatively accurate for
nanowires with $G \gsim 12 G_0$.

Ogando \etal \cite{ogando02} have self-consistently calculated
the electron shell potential for Aluminum within a local density
approximation using the stabilized jellium model. The free energy
shows oscillations characteristic of the shell effect. Despite
using a more elaborate self-consistent jellium model, their
results predict the same sequence of energetically preferred
cylindrical wires and the same positions for the supershell beat
minima as the NFEM, which \emph{a posteriori} underlines the
applicability of our approach.

Guided by the points given above, we use the NFEM to interpret
the electron-shell structure observed in \fref{fig:G22}. A
general method for performing linear stability analysis of metal
nanowires, including the modelling of material-specific surface
properties, was recently presented \cite{Urban06}. Here, we apply
the method of \onlinecite{Urban06} to Aluminum and concentrate on
presenting the results.

We examine straight wires aligned along the $z$-axis and, in order
to account for an energetically favorable symmetry breaking
(Jahn-Teller effect), both cylindrical and quadrupolar cross
sections are considered. The wire geometry is given by the radius
function
\begin{equation}
    r(\varphi)=\rho\left(\sqrt{1-\lambda^2/2}+\lambda\cos(2\varphi)\right)
\label{eq:quadrupole}
\end{equation}
in cylindrical coordinates $r,\,\varphi,\,z$, characterized by
the rms radius $\rho$ and the quadrupole deformation parameter
$\lambda$. For a given geometry $(\rho,\lambda)$ we calculate the
grand canonical potential $\Omega$ of the electron gas and
examine its change with respect to small perturbations. Only if
$\Omega$ increases for every possible small deformation of the
initial geometry is the wire (linearly) stable. Note that this
rigorous thermodynamic definition of stability is not the same as
identifying stable wires as minima of the shell potential, as is
done for example in \onlinecite{ogando02}. Although the
most-stable wires do correspond to minima of the shell potential,
not every minimum of the shell potential corresponds to a stable
wire, as has been shown by considering non-axisymmetric
deformations \cite{Urban06}.

\begin{figure}[t]       
\newlength\figonewidth\setlength\figonewidth{12.8cm}
  \begin{center}
    \includegraphics[width=\figonewidth,clip=,draft=false]{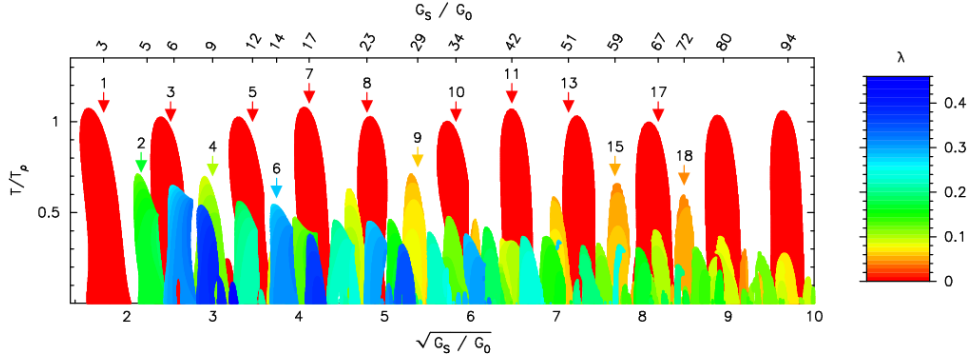}
  \end{center}
  \caption[]{(colour online) Energetically stable cylindrical and
      quadrupolar Al wires as function of temperature (in units of
      $T_\rho=T_F/k_F\rho$, see text). The deviation from axial
      symmetry is coded via the gray (colour) scale shown on the
      right. The numbered arrows label the configurations that
      enter the analysis of the shell structure and give the
      number of the corresponding peaks in the histogram.
      }
  \label{fig:StabdiaAl}
\end{figure}

\Fref{fig:StabdiaAl} displays the results of the stability
analysis as a function of temperature: shaded (coloured) areas
represent stable geometries, where the deviation from axial
symmetry $\lambda$ is coded via a gray (colour) scale. The
temperature is displayed in units of $T_\rho\equiv T_F/k_F\rho$,
where $T_F$ is the Fermi temperature, which provides a good
measure of the total stability of a metal nanowire \cite{Urban06}.

A series of very stable cylindrical wires can be identified, with
conductance values $G/G_0=3,6,12,17,23,34,42,51,...$ Their
absolute stability is periodically modulated, and remarkably
stable quadrupolar wires are found at the nodes of this supershell
structure at $G/G_0=9,29,59,...$ These two series combined form
the ``universal'' series of stable wires explaining the observed
shell and supershell structures in alkali and noble metals. Two
additional stable highly-deformed wires that are well separated in
conductance from their neighbors, and are thus expected to give
pronounced peaks in the conductance histogram, are labeled peak
Nos.\ 2 and 6 in \fref{fig:StabdiaAl}.  The theoretical quantized
conductance values and quadrupole deformations of this sequence
of stable nanowires  are given in \tref{tab:LorentzFit}.

\begin{figure}[t]     
  \begin{center}
    \includegraphics[angle=270,width=0.7\textwidth]{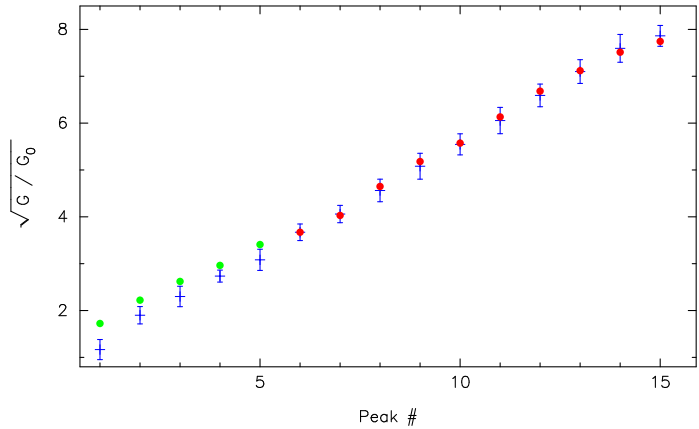}
  \end{center}
  \caption{(colour online) Comparison of the stable Al nanowires
    within the NFEM (solid circles) with the dominant experimental
    peaks (blue symbols with error bars). Here the experimental error
    bars indicate the peak widths, and the theoretical conductance
    values were corrected by a mean series resistance of 36$\Omega$.
    The NFEM predictions for the low-conductance peaks are shown in
    green, for completeness.
  \label{fig:NFEM_vs_EXP} }
\end{figure}

In \fref{fig:NFEM_vs_EXP}, this theoretical series of most stable
wires is compared to the dominant (electron-shell) peaks extracted
from the histogram of \fref{fig:G22}, plotted as (blue) $+$
symbols in \fref{fig:Periodicity}. The theoretical conductance
values have been corrected by subtracting a series resistance of
$36\,\Omega$, which was fitted to phenomenologically describe the
backscattering of electron waves by defects, surface
irregularities, and phonons \cite{Ludoph99b, Buerki99}. Aside from
a systematic shift in the theoretical peaks with $G\leq 12 G_0$,
where the NFEM is expected to be inaccurate, there is a perfect
one-to-one correspondence of the stable Al nanowires determined
within the NFEM and the experimental electron-shell structure.
This correspondence is also shown in \tref{tab:LorentzFit}, along
with the mean channel transmission $\langle T_n \rangle$ (or
series resistance $R_s$) necessary to shift the theoretical
quantized conductance values down to the experimental values. As
expected \cite{scheer97}, $\langle T_n \rangle$ is significantly
less than one for contacts with 3--6 channels due to strong mode
mixing of the $s$- and $p$-orbitals, but it approaches unity for
$G>12G_0$.  Eventually, $\langle T_n \rangle$ is expected to
decrease again as the system crosses over from ballistic to
diffusive transport.

The NFEM predicts a rather universal sequence of dominant peaks in
conductance histograms, stemming from electron-shell effects. This
sequence is expected to be present for free-electron-like metals,
and is consistent with previous experiments on alkali
\cite{yanson99,urban04} and noble metals \cite{Diaz03,mares04},
and is confirmed by the data on Aluminum reported here. For non
free-electron-like materials, other models must be used to
determine the stability. For example, classical molecular dynamics
simulations such as \cite{Pauly06}, predict very different,
material specific (i.e. nonuniversal) low conductance histograms
for Ag, Pt and Ni nanowires.

\subsection{Atomic Shell Effects}
\label{sec:AtomicShellEffects}

Similar to what is seen in aluminum clusters \cite{li98b}, the
conductance histogram in \fref{fig:G22} shows a crossover to a
different peak period of about $\Delta \sqrt{G/G_0}\simeq$ 0.25 at
larger conductance values. This shift in periodicity is associated
with the appearance of a series of peaks of increasing intensity
(cf.\ peaks A, B, C, D, 12, 14, and 16 in \fref{fig:LorentzFit})
that cannot be explained in terms of electron shell effects. This
new periodicity can be understood as an atomic shell effect: at
large diameters the nanowire will be crystalline and stable wires
are expected to form for densely-packed structures with closed
facets, that minimize the surface energy. In the case of
face-centered cubic lattice, the [110] orientation has been shown
to be most favorable to form nicely faceted long nanowires
\cite{Rodrigues00,jagla01}. We assume that the bulk fcc packing of
aluminum is preserved in nanowires similar to what has been
observed for gold and silver nanowires in transmission electron
microscopy by Rodrigues {\it et al.} \cite{Rodrigues00,
rodrigues02}. The lowest energy surfaces for Al are the $(111)$
planes \cite{vitos98} and we can construct a densely packed wire
along a $[110]$ axis with four $(111)$ facets and completing the
hexagon with two larger $(100)$ facets. One obtains a nearly
hexagonal cross-section of the wire illustrated in the inset of
\fref{fig:Al4010}. The faceting structure of stable wires is
reproduced in a Monte Carlo simulation of the thinning-down of a
fcc wire by Jagla and Tosatti \cite{jagla01}. Filling a whole
atomic shell results in an increase of the nanowire radius with
the period of $\Delta \sqrt{G/G_0}=1.98$. Assuming that stable
wires are found when completing a single facet of the hexagon, one
obtains a six times smaller period $\Delta \sqrt{G/G_0}=0.33$. An
average experimental period obtained from 11 separate measurements
would give an value of about $\Delta \sqrt{G/G_0}\sim $0.3.
 As in the case of the electronic
shell effect period, the agreement can be improved by correcting
the theoretical value by subtracting a mean series resistance of
$36\,\Omega$.

On top of this correction, a lower period is expected due to the
interplay between electronic and atomic shell effects. The
relatively higher slopes observed for samples where the atomic
shell effect is dominant, like in the histogram of
\fref{fig:Al4010}, as compared to the histogram of \fref{fig:G22}
comes in support to this assumption.

The crossover between electronic and atomic shell filling varies
between different experiments, similar to what was found for
alkali and noble metal nanowires \cite{yanson01a, mares04,
mares05} which is likely due to differences in the local crystal
orientation of the leads connecting the nanowire. On top of this,
for a given histogram the cross-over position is somewhat
arbitrary, depending on the peak-selection criteria (see
\fref{fig:Periodicity}). When considering only the dominant peaks
in the conductance histogram, mostly a value close to 40 $G_{0}$
is observed. Still, the crossover from electronic to atomic shell
effects is found to be a smooth transition rather than a sharp
feature, if we include the number of shallow ``peaks''(cf.\ peaks
A, B, C, and D in \fref{fig:LorentzFit}) which start to show up at
$G\simeq19G_0$ (cf.\ \fref{fig:Periodicity} and discussion).

These findings are in striking analogy to what was previously
found for Al clusters. The early photoionization spectrometry
measurements \cite{Schriver90,Persson91,Pellarin93} on Al$_n$
report electronic shell structure up to $n\sim100$ while
subsequent experiments \cite{Martin92,Naher93} give evidence for
atomic shell effects for $n\gsim200$. The crossover from
electronic to atomic shell effects was studied in
\onlinecite{li98b,Baguenard94} and a crossover in the range of
$n\sim75$ -- 100 atoms was reported \onlinecite{li98b}. We can
convert the number of atoms to a radius as in
\sref{sec:ElectronicShellEffects} using $v_{a}=9\pi^2/k_F^3$.
Then the crossover occurs in the range of $k_F\rho\simeq11.7$ --
12.8, which corresponds to $G/G_0\simeq28$ -- 35 via
\eref{eq:SharvinConductance}, which agrees well with our
observations.

\subsection{Superdeformed wires}
 \label{sec:Superdeformed}

\begin{figure}      
  \begin{center}
    \includegraphics[angle=270, width=0.7\textwidth]{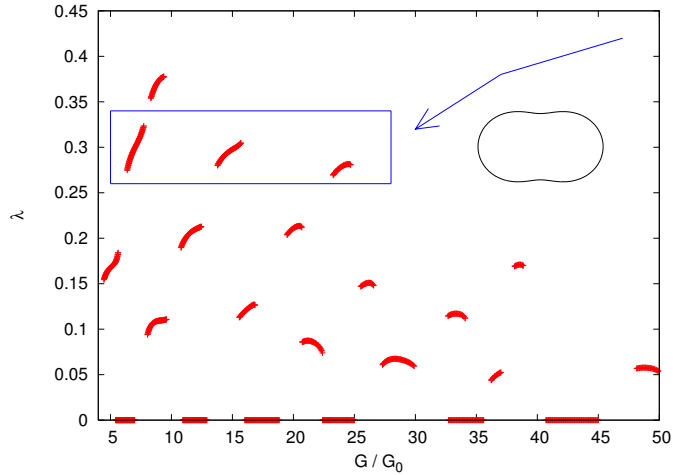}
  \end{center}
  \caption{(colour online) Stability diagram for Al wires at fixed temperature $T=0.4\,T_\rho$.
    Thick (red) lines mark stable wires in the configuration space of conductance $G$ and
    deformation parameter $\lambda$. The box emphasizes a series of very stable
    superdeformed wires, whose peanut-shaped cross section is shown as an inset.
    This sequence is thought to
    be responsible for the type of histogram shown in \fref{fig:histSuperDeformed}.}
  \label{fig:StabdiaAlphasespace}
\end{figure}

The third type of histogram seen in our experiment (cf.\
\fref{fig:histSuperDeformed}) is qualitatively different. It
shows up only at the beginning of the measurement on a fresh
sample, just after the initial breaking.  Two dominant peaks near
14 and 22 $G_0$ are observed, along with a third peak near 5
$G_0$ in two out of seven cases. An intriguing possibility is
that this third type of histogram is due to an isolated sequence
of ``superdeformed'' nanowires, that is, nanowires whose cross
section has an aspect ratio near 2:1.  Superdeformation is a
well-known feature of nuclear shell structure \cite{Nolan88}, but
has not been previously observed in nanowires.

Our stability analysis shows that the series of wires with
$\lambda\sim0.3$ ($\mbox{aspect ratio} \approx 1.9$), with
conductances of 6, 14, and 23 $G_0$, although less stable than the
magic cylinders, are quite stable compared to other
highly-deformed structures (cf.\ \fref{fig:StabdiaAl}). More
importantly, they are very isolated in configuration space, as
illustrated in \fref{fig:StabdiaAlphasespace}. Thus, if the
initial structure of the nanocontact formed in the break junction
is rather planar, with a large aspect ratio, then as the junction
is pulled apart, it will maintain a large aspect ratio as it necks
down elastically. Along such a trajectory in configuration space
(schematically indicated by a blue arrow in
\fref{fig:StabdiaAlphasespace}), the most stable structures
available are those of the superdeformed family (indicated by the
blue box).  Moreover, it is plausible that under elastic
deformation during repeated elongation/compression cycles, the
system would simply jump back and forth between these
superdeformed structures, which are geometrically similar to each
other. Eventually, however, a large structural fluctuation could
drive the nanocontact into the axisymmetric lower energy minimum,
where it would remain during further breaking cycles. This
scenario is analogous to the decay-out process of highly-excited
superdeformed nuclei, which remain in the superdeformed energy
well thoughout a long sequence of decays, before finally
tunnelling into the (lower minimum) normal-deformed energy well
\cite{Nolan88}.

Evidence of superdeformation has not been reported in any of the
previous experiments on alkali and noble metals, presumably
because highly-deformed structures are intrinsically less stable
than nearly axisymmetric structures, due to their larger surface
energy. Aluminum is unique in this respect, since its surface
tension is very small in natural units (i.e., in units of
$E_Fk_F^2$), some five times smaller than the value for
gold \cite{Tyson77}, thus favoring the possibility of large
deformations \cite{Urban06}.

\section{Summary and Conclusions}\label{sec:Summary}

We report experimental and theoretical evidence that the stability
of aluminum nanowires is governed by shell filling effects. The
experimental investigation is done by recording conductance
histograms at room temperature in UHV. Two ``magic'' series of
stable structures are observed and the corresponding periodicities
of the histogram peaks are very similar to the ones obtained
recently for noble and alkali metal nanowires. Therefore we
suggest, that also for Al the exceptionally stable structures are
related to electronic and atomic shell effects. Concerning the
electronic shell effect, the NFEM can explain the conductance and
geometry (cylindrical or quadrupolar) of the stable structures. On
the other hand, the atomic shell effect peaks can be understood
considering close-packed hexagonal structures. In addition, new
stable structures are found, that sometimes appear isolated in the
conductance histogram. Following our stability analysis, we can
attribute them to superdeformed nanowires. Remarkably, some of the
peaks having the period expected for electronic shell effect
survive under ambient conditions.

\ack We would like to thank Gijs van Dorp for assistance with the
measurements. This work is part of the research program of the
"Stichting FOM" (A.I.M. and J.M.v.R) and it was supported by the
EU Training Network DIENOW (A.I.M., D.F.U., H.G. and J.M.v.R), the
DFG (D.F.U. and H.G.) and by NSF Grant Nos.\ 0312028 (J.B.\ and
C.A.S.) and 0351964 (J.B.).

\section*{References}
\bibliographystyle{unsrt}
\bibliography{Al}


\end{document}